%% file: main.tex
\documentclass[10pt,conference]{IEEEtran}

\usepackage{cite}
\usepackage[hidelinks]{hyperref}
\usepackage{url}
\usepackage{graphicx}
\usepackage{booktabs}
\usepackage{amsmath,amssymb}
\usepackage{enumitem}
\usepackage{tikz}
\usepackage{multirow}
\usetikzlibrary{arrows.meta,positioning}
\usepackage{xcolor}
\usepackage{colortbl}
\usepackage{float}

\newcommand{\whadd}[1]{\textcolor{black}{#1}} % Additions in blue
\newcommand{\whchg}[1]{\textcolor{black}{#1}} % Changes in orange

\begin{document}

\title{GUIDE: GenAI Units In Digital Design Education}
% \author{
% \centerline{\IEEEauthorblockN{W Xiao\IEEEauthorrefmark{1},
% J Blocklove\IEEEauthorrefmark{1},
% M DeLorenzo\IEEEauthorrefmark{4},
% J Knechtel\IEEEauthorrefmark{2}, M Shafique\IEEEauthorrefmark{2},
% O Sinanoglu\IEEEauthorrefmark{2},
% K Basu\IEEEauthorrefmark{3}, 
% J Rajendran\IEEEauthorrefmark{4},
% S Garg\IEEEauthorrefmark{1},
% R Karri\IEEEauthorrefmark{1}}}
% \centerline{\IEEEauthorrefmark{1}NYU Tandon School of Engineering, New York, USA \quad
% \IEEEauthorrefmark{2}NYU Abu Dhabi, Abu Dhabi, UAE \quad
% \IEEEauthorrefmark{3}Rensselaer Polytechnic Institute, New York, USA \quad 
% \IEEEauthorrefmark{4}Texas A\&M University, Texas, USA}
% \centerline{\{wx2356, jmb9986\}@nyu.edu, matthewdelorenzo@tamu.edu,
% \{johann, ms12713, ozgursin\}@nyu.edu, basuk@rpi.edu}
% \centerline{jv.rajendran@tamu.edu, siddharth.j.garg@gmail.com, rkarri@nyu.edu}
% }

\author{
\normalsize
% W.~Xiao$^{*}$, J.~Blocklove$^{*}$, M.~DeLorenzo$^{\S}$, J.~Knechtel$^{\dagger}$,
% M.~Shafique$^{\dagger}$, O.~Sinanoglu$^{\dagger}$, K.~Basu$^{\ddagger}$,
% J.~Rajendran$^{\S}$, S.~Garg$^{*}$, R.~Karri$^{*}$\\[-0.2em]
Weihua Xiao$^{*}$, Jason Blocklove$^{*}$, Matthew DeLorenzo$^{\S}$, Johann Knechtel$^{\dagger}$, Ozgur Sinanoglu$^{\dagger}$, \\Kanad Basu$^{\ddagger}$,
Jeyavijayan Rajendran$^{\S}$, Siddharth Garg$^{*}$, Ramesh Karri$^{*}$\\
\normalsize
$^{*}$NYU Tandon, USA \quad
$^{\dagger}$NYU Abu Dhabi, UAE \quad
$^{\ddagger}$RPI, USA \quad
$^{\S}$Texas A\&M, USA\\
\normalsize
\{wx2356, jmb9986\}@nyu.edu, matthewdelorenzo@tamu.edu, \{johann, ozgursin\}@nyu.edu, basuk@rpi.edu,\\
  jv.rajendran@tamu.edu, siddharth.j.garg@gmail.com,
  rkarri@nyu.edu
\vspace{-0.5em}
}
\maketitle

\begin{abstract}

\textit{GenAI Units In Digital Design Education} (\textit{GUIDE}) is an open courseware repository with runnable Google Colab labs and other materials. 
%GUIDE organizes materials as topics, subtopics, and units.
%across LLM-aided RTL generation, LLM-aided RTL verification, and LLM-aided hardware security. 
We describe the repository's architecture and educational approach based on standardized teaching units comprising slides, short videos, runnable labs, and related papers.
%, together with recommended metadata and student deliverables. 
This organization enables consistency for both the students' learning experience and the reuse and grading by instructors.
We demonstrate GUIDE in practice with three representative units: 
% \textit{AutoChip} for tool-feedback-driven RTL generation, 
\textit{VeriThoughts} for reasoning and formal-verification-backed RTL generation, 
enhanced LLM-aided testbench generation, 
and \textit{LLMPirate} for IP Piracy. 
We also provide details for \whchg{four} example course instances (\textit{GUIDE4ChipDesign}, \textit{Build your ASIC}, \textit{GUIDE4HardwareSecurity}, and \whadd{\textit{Hardware Design}}) that assemble GUIDE units into full semester offerings, learning outcomes, and capstone
projects, all based on proven materials.
For example, the \textit{GUIDE4HardwareSecurity} course includes a project on {LLM-aided hardware Trojan insertion} that has been successfully deployed in the classroom and in \whchg{\textit{Cybersecurity Games and Conference} (\textit{CSAW})}, \whchg{a student competition and academic conference for cybersecurity}.
\whadd{We also organized an NYU Cognichip Hackathon, engaging students across 24 international teams in AI-assisted RTL design workflows.}
%a worldwide competition for cybersecurity awareness.
% We present example course instances (\textit{LLM4ChipDesign} and \textit{Build your ASIC}) that assemble GUIDE units into full semester offerings with a syllabus, learning outcomes, and capstone projects, and  a project
% on \textit{LLM-aided Hardware Trojan insertion} that has been used in the classroom and in CSAW, a world-wide  competition.
The GUIDE repository is open for contributions and available at: \url{https://github.com/FCHXWH823/LLM4ChipDesign}.
\end{abstract}

\begin{IEEEkeywords}
LLM, digital design education, RTL, hardware verification, hardware security, open-source courseware
\end{IEEEkeywords}

\input{1-Introduction}
\input{2-GUIDE}
\input{3-Submodule}
\input{4-Representative-Modules}
\input{5-LLM4ChipDesign}
\input{6-Discussion}
%\input{7-Conclusion}

% \newpage
\IEEEtriggeratref{18}
\bibliographystyle{IEEEtran}
\bibliography{Reference}
\end{document}

%% file: 1-Introduction.tex
\section{Introduction}
\label{sec:introduction}
\textit{Large Language Model}s (\textit{LLM}s)~\cite{Vaswani17} increasingly assist with key steps in digital design and verification~\cite{Xu24}, including generating \textit{Register-Transfer Level} (\textit{RTL}) code from informal specifications~\cite{Blocklove25}, producing testbenches for simulation~\cite{Bhandari25}, and translating design properties into \textit{SystemVerilog Assertion}s (\textit{SVA}s) for property checking~\cite{Xiao25}. 
This capability can lower the barrier to entry and accelerate iteration, making it attractive for education and training. Digital design comes with strict rules that make ``looks correct'' outputs risky. RTL must follow precise interfaces, clock/reset behavior, and synthesis constraints, and even small issues can lead to incorrect hardware. For instance, a missing default assignment, an unintended width truncation, or a reset mismatch can cause subtle bugs. %In addition, if the testbench is not strong enough, the design may appear correct in simulation while  being wrong.

% Teaching GenAI-based chip design introduces practical constraints beyond traditional digital design courses.
% \textbf{(1) Rapid tool evolution:} models, prompting patterns, retrieval, and finetuning practices change quickly, so ad hoc lecture material becomes obsolete in months.
% \textbf{(2) Heterogeneous cohorts:} students arrive with diverse backgrounds (UG/MS/PhD; EE/CE/CS; varying HDL and verification experience), and a fixed syllabus often fails to serve such a cohort.
% \textbf{(3) Responsible use:} LLM-assisted workflows raise concerns about IP leakage (pasting proprietary RTL into external services), evaluation contamination, and security risks (unsafe debug logic or malicious behaviors).
% \textbf{(4) Verification realism:} unlike many programming tasks, passing a small set of tests provides weak assurance for hardware; students must be trained to demand evidence.
Teaching GenAI-driven digital design raises practical challenges beyond traditional courses. First, the need for rapid tool evolution: models and prompting workflows change quickly. Course materials can be outdated within
months. Second, the need to accommodate diverse student backgrounds from different levels and majors (e.g., undergraduate vs.\ graduate; EE/CE/CS). They have different experiences with hardware description languages (HDLs) and verification. Hence,
one syllabus may not fit everyone. Third, the need for use of complex tools: LLM-aided digital design is not a standalone step but an end-to-end workflow. This workflow must be integrated with EDA tools for compilation, simulation, and verification and this integration needs to be taught.

We introduce \textit{GUIDE}, \textit{GenAI Units In Digital Design Education} to teach LLM-aided digital design. The key idea is to make course content reusable and easy to update. We build a public repository with hands-on Google Colab labs. 
%The repository is organized as topics $\rightarrow$ subtopics $\rightarrow$ units. 
Each unit is a teaching unit with a clear scope. It comes with slides, a short video, a runnable lab, and related paper. 
% Each unit also includes recommended metadata that helps instructors select suitable units for a course.
% We also show how new units can be added over time. 
% Finally, we present \textbf{\textit{GenAI4ChipDesign}} as an example course built from the repository. We summarize how it selects and orders modules. We also include its syllabus, learning outcomes, and a capstone project, {LLM-Based  Verification}.
Finally, we present \whchg{four} course instances built from the repository, \textit{GUIDE4ChipDesign}, \textit{Build your ASIC}, \textit{GUIDE4HardwareSecurity}, and \whadd{\textit{Hardware Design}}, and summarize how we selected and organized units into a semester offering, including a syllabus  and a capstone. 
% We include a project on \textit{LLM-aided Trojan insertion} for classroom teaching and  competitions.

%In the remainder of this paper, This paper makes five contributions. 
In Section~\ref{sec:GUIDE}, we outline the GUIDE architecture and summarize the main topic coverage in our repository. 
%In Section~\ref{sec:sub-unit},
Furthermore, we define what a unit is and provide a standard unit architecture, including required teaching materials, and quality requirements.
%, recommended metadata, and guidelines for adding new units. 
In Section~\ref{sec:representative-sub-units}, we present three sample units to show what a teaching-ready unit looks like in practice. 
In Section~\ref{sec:course-instances}, we report \whchg{four} example courses built from GUIDE.
In Section~\ref{sec:discussion}, we discuss  future directions.

%% file: 2-GUIDE.tex
\section{GUIDE Architecture}
\label{sec:GUIDE}
\begin{figure}[tb]
\centering
%\hspace{-0.07\columnwidth}\includegraphics[width=1.07\columnwidth]{fig/overview_new.png}
\includegraphics[width=1.07\columnwidth]{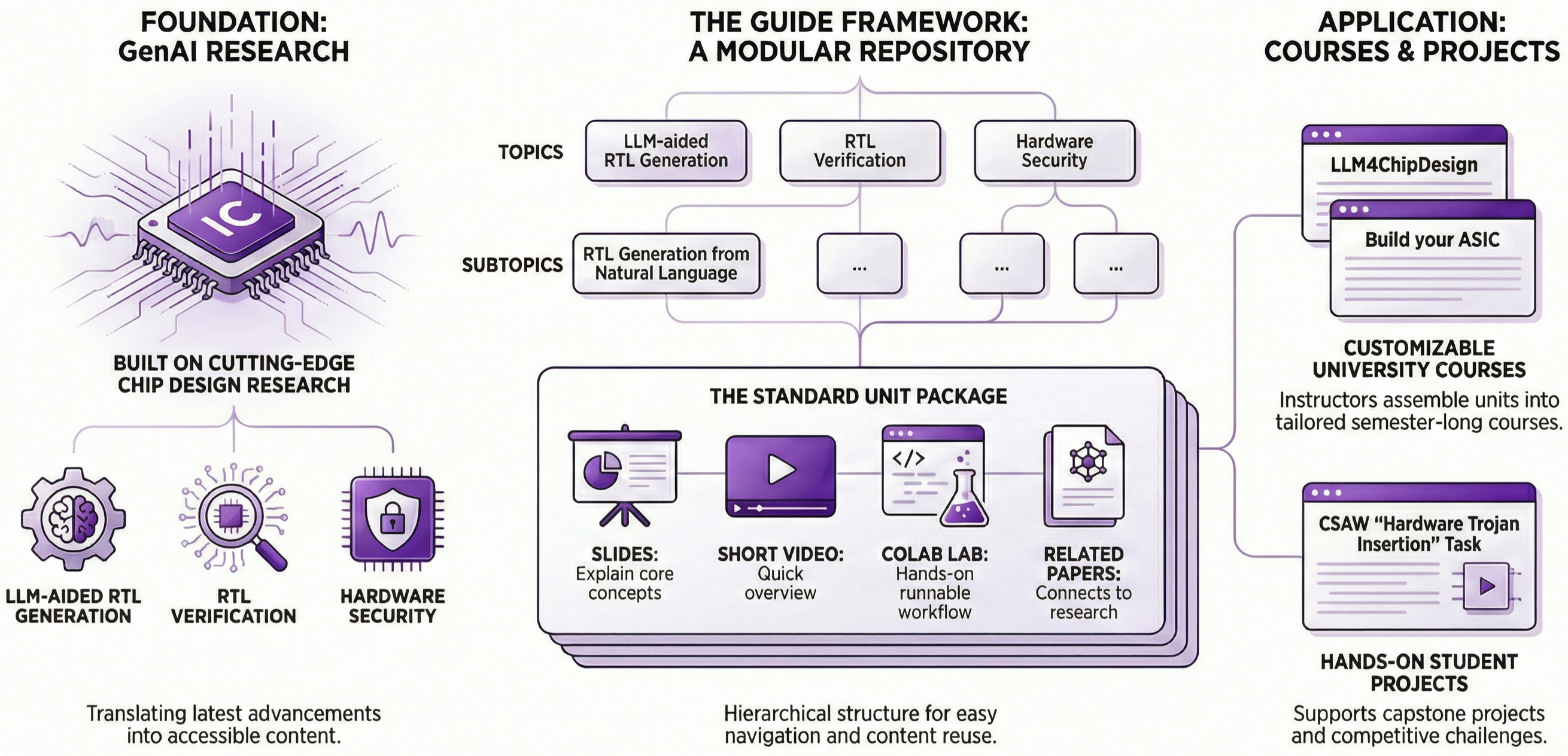}
\vspace{-4mm}
\caption{Educational approach for generative AI in chip design using the open-source GUIDE framework. Figure was generated using NotebookLM.}
\label{fig:overview} 
\vspace{-4mm}
\end{figure}

%In this paper, 
%\textbf{\textit{GUIDE}} stands for 
GUIDE 
(see Fig.~\ref{fig:overview}) 
provides an educational approach and an open-source courseware repository for instructors to teach GenAI topics in digital design in a structured and reusable way. The goal is to turn research results into material that can be used in classes. The material should be easy to adopt, easy to update, and suitable for students with different backgrounds.

% GUIDE has a modular structure as shown in Table~\ref{tab:guide_taxonomy}. 
% The repository is organized into {topics}, {subtopics}, and {units}, in that hierarchical order.
% GUIDE is an open and scalable repository currently focused on three topics: {LLM-aided RTL generation}, {LLM-aided RTL verification}, and {LLM-aided hardware security}. 
% Each topic has several subtopics.
% For RTL generation, this includes RTL generation from \whadd{natural language (\textit{NL})} and finetuned LLMs.
% For RTL verification, this includes LLMs for simulation-based
% verification and LLMs for formal verification. 
% For hardware security, this includes LLMs for attacks and defenses. 
% Each subtopic has different units.
% A unit is the smallest teaching-ready element and corresponds to one concrete teaching objective.
GUIDE has a modular structure (Table~\ref{tab:guide_taxonomy}) organized into {topics}, {subtopics}, and {units}.
Its three topics are: \textit{LLM-aided RTL generation} (subtopics: generation from \whadd{natural language (\textit{NL})}, finetuned LLMs), \textit{LLM-aided RTL verification} (subtopics: simulation-based and formal verification), and \textit{LLM-aided hardware security} (subtopics: attacks and defenses).
% Each subtopic has different units.
A unit is the smallest teaching-ready element, corresponding to one concrete teaching objective. 

%% file: 3-Submodule.tex
\subsection{Standard Unit Structure}
GUIDE uses a uniform unit structure so instructors can teach with low setup cost and students can use it easily. Each unit is 
%(Table~\ref{tab:submodule_package}). These 
simple and ``teaching-ready'' and has four elements:
\begin{itemize}[leftmargin=*]
  \item 
  \textbf{Slides} present the core idea of the unit. They should introduce the problem, define the inputs/outputs, and explain the key concepts that students need. Slides should highlight common pitfalls and include an  example.
  \item 
  \textbf{Short video} provides a quick overview for self-study and review. It should explain what the unit does, the workflow, and what results students should expect.
  \item 
  \textbf{Colabs}  provide a hands-on workflow to run end-to-end (setup, run, check). It includes at least one runnable example so students can execute the main steps and see the expected artifacts. If Google Colab is
  not applicable (e.g., special tools, licensing limits, or heavy compute), the unit should provide a runnable script and clear instructions.
  \item 
  \textbf{Related papers}  connect the unit to research context and deeper reading. This helps instructors justify the unit and motivates students \whchg{to} explore beyond the lab.
\end{itemize}

%\begin{table}[t]
%\centering
%\caption{Required materials for each unit.}
%\label{tab:submodule_package}
%\setlength{\tabcolsep}{6pt}
%\renewcommand{\arraystretch}{1.15}
%\begin{tabular}{|p{1.7cm}|p{6.0cm}|}
%\hline
%\rowcolor{gray!20}
%\textbf{Item} & \textbf{Purpose} \\
%\hline\hline
%Slides & Problem definition, key ideas, and common pitfalls. \\
%Short video & Quick overview for self-study and review. \\
%Colab lab & Workflow (setup, run, check) with an example. \\
%Related papers & Background reading and research context. \\
%\hline
%\end{tabular}
%\vspace{-5pt}
%\end{table}

\subsection{Quality Requirements for Teaching and Reuse}
A unit should be ``teaching-ready'' and easy to reuse across different courses. We use the following requirements.
\begin{itemize}[leftmargin=*]
  \item \textbf{Runnable from scratch}: The lab should run in a clean environment. It should not depend on private paths or local files. If extra packages are needed, the lab should install them or list them clearly.

  \item \textbf{End-to-end runnable example}: Each unit should include at least one runnable example that students can execute end-to-end. The example should make the workflow concrete and show the expected artifacts.

  \item \textbf{Evidence for grading}: The unit should specify what students must submit as evidence, such as compilation/ simulation logs, waveform screenshots, or reports. This reduces the risk of solutions that ``look correct.''

  \item \textbf{Stable interfaces and file layout}: If a unit generates RTL, testbenches, or assertions, it should follow a consistent file naming and folder layout. This makes it easier to combine units into a  course project.

  \item \textbf{Clear explanations}: Slides and labs should use simple language. Advanced terms should be defined.
\end{itemize}

% \subsection{Recommended Metadata}
% To help instructors select suitable units, each one should provide metadata as a short text file or as a JSON/YAML:
% \begin{itemize}[leftmargin=*]
%   \item \textbf{Topic and subtopic:} where the unit belongs.
%   \item \textbf{Difficulty level:} introductory, intermediate, or advanced.
%   \item \textbf{Prerequisites:} required background knowledge and tools.
%   \item \textbf{Expected time:} typical hours needed by students.
%   \item \textbf{Required tools:} simulator, assertion checker etc.
%   \item \textbf{Expected outputs:} generated artifacts such as RTL, testbenches, assertions, logs, or reports.
%   \item \textbf{Evidence:} what students should provide for grading.
% \end{itemize}
% This metadata supports building courses from the same repository and helps students understand the setup and workload.

% \subsection{Integration of New Units}
% Adding a new unit should be straightforward. A contributor prepares the four required materials in Table~\ref{tab:submodule_package}. The contributor also adds the metadata described above. Finally, the contributor verifies that the lab runs end-to-end from a clean setup and records what evidence students should submit. These steps keep the repository consistent as it grows.
% More details are available in our GitHub repository at \url{https://github.com/FCHXWH823/LLM4ChipDesign}.

\begin{table*}[tbp]
\centering
\vspace{-15pt}
\caption{GUIDE taxonomy in our current repository: topics, subtopics, and representative units and descriptions.}
\label{tab:guide_taxonomy}
\setlength{\tabcolsep}{6pt}
\renewcommand{\arraystretch}{1.25}
\footnotesize
\begin{tabular}{|m{1.6cm}|m{2.4cm}|m{3.2cm}|m{9.4cm}|}
\hline
\rowcolor{gray!20}
\centering\textbf{Topic} & \centering\textbf{Subtopic} & \centering\textbf{Unit} & \centering\arraybackslash\textbf{Description} \\
\hline\hline
% ===== Topic 1: LLM-aided RTL generation =====
\multirow{19}{=}[10pt]{\centering\textbf{LLM-aided RTL Generation}}
& 
\multirow{10}{=}[10pt]{\centering RTL Generation from NL}
& \centering AutoChip~\whchg{\cite{thakur2023autochip}}
& Generate Verilog from a prompt and testbench plus iterative compilation/ simulation feedback. \\
\cline{3-4}
& & \centering ROME~\cite{Nakkab24}
& Uses hierarchical prompting to decompose complex designs so smaller open-source LLMs can generate larger Verilog systems with better quality and lower cost. \\
\cline{3-4}
& & \centering Veritas~\cite{Roy25}
& Has an LLM generate CNF clauses as a formal functional specification and deterministically converts CNF to Verilog for correctness by construction. \\
\cline{3-4}
& & \centering PrefixLLM~\cite{Xiao24}
& Represents prefix-adder synthesis as structured text (SPCR) and performs iterative LLM-guided design space exploration to optimize area and delay. \\
\cline{3-4}
& & \centering VeriDispatcher~\cite{Wang25}
& Dispatch RTL tasks to LLMs using pre-inference difficulty prediction to improve quality and reduce LLM use cost. \\
\cline{2-4}
& \multirow{10}{=}[10pt]{\centering Finetuned LLMs for RTL Generation}
& \centering \whchg{VGen}~\whchg{\cite{Thakur23}} & \whchg{Proposes a dedicated dataset and demonstrates that finetuning LLMs on this dataset significantly improves their Verilog code generation capabilities.}
\\
\cline{3-4}
& & \centering VeriThoughts~\cite{Yubeaton25}
& Provides a formal-verification-based pipeline to build a reasoning-oriented Verilog dataset and to finetune LLMs for high-accuracy Verilog generation. \\
\cline{3-4}
& & \centering VeriReason~\cite{VeriReason25}
& A DeepSeek-R1-inspired RTL generation framework that combines supervised finetuning with GRPO reinforcement learning and feedback-driven rewards. \\
\cline{3-4}
& & \centering VeriContaminated~\cite{wang2025vericontaminated}
& Analyzes data contamination in Verilog benchmarks (VerilogEval, RTLLM) to assess the validity and fairness of SOTA LLM code generation evaluations. \\
\cline{3-4}
& & \centering LLM Architecture Insights~\cite{karn2026ISCAS}
& Inspects LLM internals (layers, attention, context) and visualizes parameter shifts during finetuning to understand suitability for hardware constraints. \\
\hline
% ===== Topic 2: LLM-aided validation and verification =====
\multirow{9}{=}[10pt]{\centering\textbf{LLM-aided RTL Verification}}
& \multirow{5}{=}[10pt]{\centering Simulation-based Verification}
& \centering Testbench Generation~\cite{Bhandari25}
& Given the RTL under test and an NL description of the golden RTL, it generates comprehensive test patterns and then refines them using feedback from EDA tools to improve coverage and expose bugs. \\
\cline{3-4}
& & \centering Enhanced Testbench Generation
& Starting from the RTL  and its description, generate a testbench, build a Python golden model to compute outputs, insert self-checking logic, and run simulation. \\
\cline{2-4}
& \multirow{7}{=}[10pt]{\centering  Formal Verification}
& \centering RAG-based SVA Generation
& Builds a knowledge base from OpenTitan documentation and uses retrieval-augmented generation to produce context-aware SVAs for IP blocks. \\
\cline{3-4}
& &\centering SV Assertions \cite{menon2025enhancing}
& Utilizing LLMs to generate SystemVerilog assertions from design documentation. \\
\cline{3-4}
& &\centering Assert-O \cite{miftah2024assert}
& Optimization of SystemVerilog assertions using LLMs. \\
\cline{3-4}
& & \centering Hybrid-NL2SVA~\cite{Xiao25}
& A RAG framework for NL2SVA and a finetuning pipeline with a synthetic dataset to train lightweight LLMs to translate NL properties into SVAs. \\
\hline
% ===== Topic 3: LLM for hardware security =====
\multirow{17}{=}[10pt]{\centering\textbf{LLM-aided Hardware Security}}
& \multirow{9}{=}[10pt]{\centering Hardware Attacks}
& \centering LLMPirate~\cite{Gohil24}
& LLM-driven rewriting to thwart piracy-detection tools. \\
\cline{3-4}
% & & \centering TrojanPlayground~\cite{Sarihi22} & Automatically insert stealthy combinational hardware Trojans by exploring circuit design spaces to maximize concealment using reinforcement learning. \\
% \cline{3-4}
& & \centering ATTRITION~\cite{Gohil22} & An RL framework that models an adversary to  evaluate and evade prior Trojan detectors, showing dramatically higher attack success than random insertion. \\
\cline{3-4}
& & \centering \whchg{GHOST}~\whchg{\cite{Faruque24}} & \whchg{An automated LLM-based attack framework that generates and inserts stealthy, synthesizable Hardware Trojans into HDL designs, enabling rapid Trojan creation and highlighting detection risks in modern hardware security flows.} \\
\cline{3-4}
%& & \centering NetDeTox~\cite{wang2025netdetox}
%& Orchestrates RL and LLMs to perform adversarial netlist rewrites that evade GNN-based hardware security detection with minimal overhead. \\
\cline{3-4}
& & \centering RTL-Breaker~\cite{mankali2025rtlbreaker}
& A framework assessing backdoor attacks on LLM-based HDL generation, analyzing trigger mechanisms and their impact on code quality and security. \\
\cline{3-4}
%& & \centering Netlist Whisperer~\cite{roy2025netlist}
%& Uses finetuned LLMs and ensemble methods to analyze netlists and identify side-channel leakage vulnerabilities without full simulation. \\
\cline{2-4}
& \multirow{12}{=}[10pt]{\centering Hardware Defenses}
& \centering Security Assertions~\cite{Kande24}
& LLM-generated security assertions from NL prompts/comments. \\
\cline{3-4}
& & \centering NOODLE~\cite{Vishwakarma24} & A multimodal, risk-aware Trojan detection unit that addresses limited Trojan benchmarks by using GAN-based data augmentation and a multimodal deep learning detector with uncertainty estimates for decision making. \\
\cline{3-4}
& &\centering \whchg{NSPG} \cite{meng2024nspg}
& \whchg{Security property generator based on natural-language processing (NLP).} \\
\cline{3-4}
% & & \centering TrojanSAINT~\cite{Lashen23} & A graph neural network-based Hardware Trojan detection scheme working at the gate level. \\
% \cline{3-4}
& & \centering TrojanLoC~\cite{XiaoTrojanLoC25}
& Use RTL-finetuned LLM embeddings plus lightweight classifiers to detect Trojans, predict types, and localize suspicious lines using TrojanInS dataset. \\
 \cline{3-4}
& & \centering LockForge~\cite{saha2025lockforge}
& A multi-agent LLM framework that automates the translation of logic locking schemes from research papers into executable, validated code. \\
\cline{3-4}
& & \centering SALAD~\cite{wang2025salad}
& An  assessment framework using machine unlearning to remove sensitive IP, contaminated benchmarks, and malicious patterns from LLMs without full retraining. \\
%& & \centering VeriLeaky~\cite{wang2025verileaky}
%& Investigates IP leakage risks when finetuning LLMs on proprietary Verilog and evaluates logic locking as a mitigation strategy to balance utility and protection. \\
\cline{3-4}
\hline
\end{tabular}
\vspace{-5pt}
\end{table*}

%% file: 4-Representative-Modules.tex
\section{Representative Units}
\label{sec:representative-sub-units}
This section describes representative units in the GUIDE repository.
Each example follows a consistent format. We summarize the learning goal, key student outputs, and the required checks. These examples show how the
repository covers multiple topics in GenAI-driven digital design education.

% \subsection{AutoChip: Tool-Feedback-Driven RTL Generation}
% \label{subsec:autochip}

% \subsubsection{Goal}
% AutoChip teaches students how an LLM can generate RTL code from an NLP specification, and how to improve correctness by iterating with EDA tools (compilation and simulation) instead of relying on code that ``looks
% correct''.

% \subsubsection{Workflow}
% Students provide an English specification, a Verilog module interface, and a testbench. AutoChip queries an LLM to generate candidate implementations, then compiles and simulates them with the testbench. If
% compilation fails or simulation mismatches occur, AutoChip converts the tool outputs into a feedback prompt and re-queries the LLM to repair the current code. 
% The loop repeats until the RTL code both compiles and passes simulation, or a maximum iteration budget is reached. 
% The lab discusses practical choices for keeping feedback concise versus full history, and an optional small-to-big model handoff when a lightweight model fails.

% \subsubsection{Outputs and Submission}
% Students submit the RTL, the prompts (initial design prompt plus at least one feedback prompt), and an iteration note describing what failed first (compile or simulation) and what changed across iterations to fix it.
% %\subsubsection{ Submission}
% Students submit compilation/ simulation log showing a pass result, plus one example of an earlier failure (a compile error or simulation mismatch) and corresponding repair.

\subsection{VeriThoughts: Reasoning and Formal Verification}

\subsubsection{Goal}
\whchg{This unit teaches students to understand how verification-backed reasoning datasets are constructed and to analyze RTL generation quality through the lens of formal equivalence checking.}
Students learn why NL prompts alone can be noisy, and how a formal-equivalence-checked synthesis pipeline produces more reliable training data.
The unit connects course topics (prompting, RTL generation, and formal verification) through a hands-on end-to-end example. % based on the VeriThoughts pipeline.

\subsubsection{Workflow}
\whchg{Students first select a subset of ground-truth Verilog modules from the VeriThoughts dataset. They then run the VeriThoughts synthesis pipeline on this subset: an LLM generates an NL question for each module, a reasoning-capable LLM produces a step-by-step reasoning trace and a candidate Verilog solution, and a formal equivalence checker compares each candidate against the ground-truth design and labels it as match or mismatch. Students then analyze the synthesis results --- for example, examining how reasoning trace length, prompt style, or module complexity correlates with formal equivalence outcomes.}
%\whadd{If GPU resources are available (e.g., a GPU-enabled Colab Pro environment or a departmental cluster), students may additionally finetune a lightweight model on the synthesized data.}

\subsubsection{Outputs and Submission}
\whchg{Students submit: (i) the selected subset configuration (which modules, what filtering criteria), (ii) the synthesized dataset including generated NL questions, reasoning traces, candidate Verilog solutions, and formal equivalence labels, (iii) a formal equivalence checker log for at least one match and one mismatch example, and (iv) a short analysis report discussing how pipeline or subset choices affect the match/mismatch distribution.}
%\whadd{If GPU resources are available, students additionally submit a finetuned model checkpoint or reproducible training script with loss curves and benchmark evaluation results.}

\subsection{Enhanced LLM-Aided Testbench Generation}

\subsubsection{Goal}
This unit teaches students how to build an end-to-end testbench generation workflow that produces a self-checking simulation result. The key idea is to pair LLM-generated test patterns with an independent ``golden'' reference model so the final testbench can automatically report pass/fail.

\subsubsection{Workflow}
The lab follows a 5-step pipeline: (1--2) take the NL description and the Verilog design-under-test as inputs, (3) use an LLM to generate test patterns and a testbench skeleton, (4) generate a Python golden model from
the NL description and compute expected outputs, and (5) enhance the testbench with expected outputs and self-checking logic.
%with a clear test summary.

\subsubsection{Outputs and Submission}
Students include a testbench with test patterns, a Python golden reference implementation, a pattern file with expected outputs, and a final self-checking testbench that compares outputs against golden outputs and reports pass/fail.
%\subsubsection{Evidence to submit}
Students also include artifacts and a simulation log showing compilation and the final pass/fail summary produced by a standard RTL simulation flow. This is easy to grade while reflecting whether the workflow runs end-to-end.

\subsection{LLMPirate: LLMs for IP Piracy}
\subsubsection{Goal} LLMPirate teaches students how LLMs can be maliciously leveraged to obfuscate a circuit design to evade piracy detection tools while maintaining functionality. Students will learn the initial limitations of LLMs in rewriting large-scale Verilog code, and develop a prompting framework to overcome context-window and training challenges to successfully pirate circuit designs from an attacker's perspective.
\subsubsection{Workflow}
This lab consists of developing an iterative prompting framework for hardware IP piracy. The assignment begins with students directly prompting an LLM to rewrite a set of Verilog circuits such that the gate-level
structure is different, while maintaining functionality. Then, the generated circuits will be evaluated for functional equivalence and structural similarity to their original counterparts through formal-equivalence and
piracy-detection tools. Students will utilize a Boolean representation of the circuit netlists within the prompt, instructing the model to perform gate-level transformations. Then, students will enhance the framework
through defining iterative feedback prompts that utilize the output of the provided tools. The circuits will be evaluated for functionality and piracy evasion after each implementation.
\subsubsection{Outputs and Submission}
Students will provide (i) their final prompting framework consisting of the initial prompt and all tool-feedback prompts, (ii) the functional equivalence and structural similarity scores for all tested Verilog circuits,
and (iii) an evaluation of scores after each framework implementation, assessing which configuration best facilitated IP piracy. 
% \subsection{Summary and coverage}
% Together, these units cover three current directions in GUIDE: (i) LLM-aided RTL generation, (ii) LLM-aided validation and verification (testbenches and SVAs), and (iii) security-aware digital design learning. Additional units can be added over time under the same format. This makes it possible to expand the repository to new topics such as LLM-aided high-level synthesis (HLS) and code generation for analog/SPICE workflows.

%% file: 5-LLM4ChipDesign.tex
\section{Examples for GUIDE-Driven Courses}
\label{sec:course-instances}

GUIDE is not tied to a single syllabus. Instead, instructors can select GUIDE units and
organize them into course instances that match their audience, timeline, and learning goals.
We present \whchg{four} examples next.

% ----------------------------
% =========================
% Section V-A and V-B (LaTeX)
% =========================
% Section V: Example GUIDE Course Instances

\begin{table*}[t]
%\vspace{-8pt}
\centering
\caption{Semester plan for GUIDE4ChipDesign (week-level view).}
\vspace{-5pt}
\label{tab:plan_llm4chipdesign}
\renewcommand{\arraystretch}{1.12}
\begin{tabular}{|c|p{3.2cm}|p{12.2cm}|}
\hline
\rowcolor{gray!20}
\textbf{Week} & \centering \textbf{Theme} \arraybackslash & \centering \textbf{Typical Activities} \arraybackslash \\
\hline \hline
1 & Course setup & Course overview; GenAI for digital design; environment setup; first RTL-generation warm-up lab. \\
\hline
2 & Prompting basics & NL specification writing; interface constraints; compile-first mindset. \\
\hline
3 & RTL generation I & Generate small combinational Verilog modules; fix compile errors; learn common syntax pitfalls. \\
\hline
\multirow{2}{*}{4} & \multirow{2}{*}{RTL generation II} & Sequential RTL patterns; clock/reset conventions; debug reset mismatches. Analyzing fine-tuning dynamics: visualizing parameter shifts in domain-adapted models. \\
\hline
5 & Debugging practice & Iterative refinement using compiler/simulator feedback; style and structure checks. \\
\hline
6 & Testbenches I & Self-checking testbenches; directed tests; pass/fail reporting. \\
\hline
7 & Testbenches II & Corner cases and boundary values; debugging using waveforms and logs. \\
\hline
8 & Enhanced validation & Stronger test patterns; add a simple reference model when possible; evidence-based submissions. \\
\hline
9 & Properties I & What properties mean; mapping intent to SVAs; basic temporal operators. \\
\hline
10 & Properties II & Common SVA pitfalls; vacuity; disable conditions; review and correction using examples. \\
\hline
11 & NL2SVA practice & Translate design properties into SVAs; compile checks; small-scale trace reasoning when needed. \\
\hline
12 & Security awareness & Security mindset for RTL; audit thinking; suspicious patterns to look for. \\
\hline
% 13 & Security awareness II & Simple negative tests; validate unexpected behaviors via simulation evidence. \\
% \hline
13 & Project integration & Integrate RTL + testbench + checks; project milestone review and debugging clinic. \\
\hline
14 & Capstone wrap-up & Final integration, results packaging, and presentation of evidence (logs/outputs). \\
\hline
\end{tabular}
%\vspace{-5pt}
\end{table*}

\begin{table*}[htbp]
\centering
\caption{Semester plan for Build your ASIC (week-level view).}
\vspace{-5pt}
\label{tab:plan_build_your_asic}
\renewcommand{\arraystretch}{1.12}
\begin{tabular}{|c|p{5.2cm}|p{10.3cm}|}
\hline
\rowcolor{gray!20}
\textbf{Week} & \centering \textbf{Theme} \arraybackslash & \centering \textbf{Typical Activities} \arraybackslash \\
\hline \hline
1 & Introduction & Course overview; setup; expectations and tooling. \\
\hline
2 & Combinational logic + testbench & Basic modeling; testbench structure; initial lab work. \\
\hline
3 & Combinational logic + testbench & Continue Week 2; submit Week 2 work. \\
\hline
4 & Sequential logic modeling 1 + testbench & Sequential basics; submit Week 3 work. \\
\hline
5 & Sequential logic modeling 2 + testbench & Submit Week 4 work; preliminary project proposal. \\
\hline
6 & LLM-aided chip design 1 & Use LLMs to design combinational/sequential logic and testbench; submit Week 5 work. \\
\hline
7 & LLM-aided chip design 2 & Continue LLM-aided workflow; finalize project plan. \\
\hline
8 & Project simulation (combinational) & Run simulation for the combinational part of the project. \\
\hline
9 & Project simulation (sequential) & Simulate sequential part; submit to TinyTapeout. \\
\hline
10 & TinyTapeout submission & Submit final project to TinyTapeout workflow. \\
\hline
11 & Documentation + presentations & Present progress; prepare chip tapeout documentation. \\
\hline
12 & Project report & Draft the project report and collect workflow artifacts. \\
\hline
13 & Final report submission & Submit final report and required artifacts. \\
\hline
\whadd{14} & \whadd{Course wrap-up} & \whadd{Final presentations, feedback, and course review.} \\
\hline
\end{tabular}
%\vspace{-5pt}
\end{table*}

\begin{table*}[htb]
\vspace{-8pt}
\centering
\caption{Semester plan for GUIDE4HardwareSecurity (week-level view).}
\vspace{-5pt}
\label{tab:llm4hwsec_plan}
\setlength{\tabcolsep}{6pt}
\renewcommand{\arraystretch}{1.12}
\footnotesize
\begin{tabular}{|c|p{4.5cm}|p{11.5cm}|}
\hline
\rowcolor{gray!20}
\textbf{Week} & \centering \textbf{Theme} & \centering \textbf{Typical Activities} \arraybackslash \\
\hline\hline
1 & Course setup & Course overview; GenAI for digital design; environment setup; first RTL-generation warm-up lab. \\
\hline
2 & Prompting basics & NL specification writing; interface constraints; compile-first mindset. \\
\hline
3 & RTL generation I & Generate small combinational Verilog modules; fix compile errors; learn common syntax pitfalls. \\
\hline
\multirow{2}{*}{4} & \multirow{2}{*}{RTL generation II} & Sequential RTL patterns; clock/reset conventions; debug reset mismatches. Analyzing fine-tuning dynamics: visualizing parameter shifts in domain-adapted models. \\
\hline
5 & Security foundations & Introduction to hardware security threats; Trojan taxonomy; attack surfaces in digital design. \\
\hline
6 & LLM-aided attacks I & LLMPirate: LLM-driven IP piracy and obfuscation techniques; evading detection tools. \\
\hline
7 & LLM-aided attacks II & GHOST: Automated Trojan insertion framework; generating stealthy, synthesizable Trojans. \\
\hline
8 & Advanced attacks & ATTRITION: RL-based adversarial framework for hardware designs. \\
\hline
9 & Defense foundations & Security assertions and property checking; translating security requirements to verifiable properties. \\
\hline
10 & LLM-aided defenses I & Security Assertions: generating security properties from NL prompts; validation. \\
\hline
11 & LLM-aided defenses II & TrojanLoC: Trojan detection and localization using LLM embeddings; TrojanInS dataset exploration. \\
\hline
12 & Advanced defenses & NOODLE: multimodal Trojan detection; LockForge: automated logic locking implementation. \\
\hline
13 & Attack-defense integration & SALAD: machine unlearning for sensitive IP protection; integrated attack-defense workflows. \\
\hline
14 & Capstone presentations & Team presentations of integrated Trojan insertion-detection projects; peer evaluation and discussion. \\
\hline
\end{tabular}
\end{table*}

\begin{table*}[htbp]
\centering
\caption{Semester plan for Hardware Design (week-level view).}
\vspace{-5pt}
\label{tab:hardwareDesign}
\renewcommand{\arraystretch}{1.12}
\begin{tabular}{|c|p{6cm}|p{9cm}|}
\hline
\rowcolor{gray!20}
\textbf{Week} & \centering \textbf{Theme} \arraybackslash & \centering \textbf{Typical Activities} \arraybackslash \\
\hline \hline
1 & Introduction & Overview, tool setup, and first Verilog RTL lab. \\
\hline
2 & Combinational blocks + testbench & Encoders/decoders, mux design, and ALU testbench lab. \\
\hline
3 & Sequential logic elements & Flip-flops/latches with traffic-light controller lab. \\
\hline
4 & Sequential circuit modeling+analysis & FSM/state-table modeling and traffic-light verification lab. \\
\hline
5 & Sequential circuit design 1 & Spec-to-circuit design with counter and pattern-detector labs. \\
\hline
6 & Sequential circuit design 2 & Registers, hierarchical Verilog, and CPU design lab. \\
\hline
7 & Sequential circuit design 3 & Counters, PLAs, memory hierarchy, and CPU testbench lab. \\
\hline
8 & LLM-aided chip design 1 & LLM-based RTL generation and advanced CPU coding lab. \\
\hline
9 & LLM-aided chip design 2 & LLM RTL generation for neural-accelerator modules and tests. \\
\hline
10 & LLM-aided RTL verification 1 & LLM-assisted testbench generation for neural-accelerator blocks. \\
\hline
11 & LLM-aided RTL verification 2 & Enhanced validation and RAG-based SVA generation lab. \\
\hline
12 & Project: LLM-aided core generation and validation 1 & MAC-unit generation, validation, and simulation milestone. \\
\hline
13 & Project: LLM-aided core generation and validation 2 & MAC-array design/validation and integrated simulation milestone. \\
\hline
14 & Project presentation and course wrapup & Final report, presentations, feedback, and course wrap-up. \\
\hline
\end{tabular}
%\vspace{-5pt}
\end{table*}

\subsection{Example Course 1: GUIDE4ChipDesign I \& II}

\subsubsection{GUIDE4ChipDesign I}
This part targets students who already know basic digital design and Verilog. The course focuses on how to use LLMs in  digital-design workflows, including RTL generation, simulation- and assertion-based verification.
The course had 27 students enrolled.

\textbf{GUIDE Units:} GUIDE4ChipDesign I selects units spanning three GUIDE topics:  LLM-aided (i) RTL generation, (ii)  RTL verification, and (iii) hardware security.

\textbf{Semester Plan:} See Table~\ref{tab:plan_llm4chipdesign}. Early weeks focus on writing clear specifications, debugging RTL codes with tool feedback, and architectural insights for LLM-aided hardware design.
Middle weeks focus on testbench generation. Later weeks introduce properties and SVAs. Instructors can swap or reorder weeks based on student background and course goals. 

\textbf{Capstone Project:} The final project is \textit{LLM-Based Verilog Adder Generation and Verification}. This capstone guides students through an end-to-end LLM-aided digital design workflow using adders as a
focused case study. Students select two different adder architectures from a public repository of golden implementations, reverse engineer each design into a detailed NL description (architecture, hierarchy, and signal behavior), and use an LLM tool (e.g., AutoChip in our GUIDE repository) to regenerate Verilog that follows the description and matches the required interface. Students then compare the regenerated RTL code against the golden reference at a high level (e.g., module structure and key signals). Next, students use an enhanced LLM-aided testbench generator to produce self-checking testbenches that validate both primary outputs and selected internal signals, and they run RTL compilation and simulation to report pass/fail evidence.

% \textbf{Assessment and deliverables:} Assessment combines homework labs and the capstone project. Labs emphasize runnable artifacts and clear evidence, such as compile logs, simulation logs, and (when used) SVA compilation or property-checking outputs. Grading rewards correctness evidence and clear reporting, not only code length.

\subsubsection{GUIDE4ChipDesign II}
This builds on the first semester and transitions to team-based design projects. The course had 13 teams (two students per team) who propose their own projects. Teams apply the LLM-aided design, verification, and
security techniques from GUIDE4ChipDesign I to complete a full implementation including FPGA deployment. The semester follows a milestone-based structure with weekly presentations.
%Each team submits: (1) weekly progress presentations, (2) final project report (5--6 pages minimum) with design details, LLM tool usage, simulation and synthesis results, and FPGA validation evidence, (3) GitHub repository with complete source code and LLM interaction logs, (4) demonstration video (10 minutes) showing the project working on FPGA board with test/corner cases.
Other deliverables are a final report (design logs, FPGA validation), a GitHub repo, and a demo video.

\whadd{\textbf{NYU Cognichip Hackathon.}
We organized a hands-on activity that exposed 72 students across 24 teams (21 from the US, one from Canada, two from India) to AI-assisted design workflows.
Students used the \textit{Cognichip}~\cite{cognichip} platform to develop RTL solutions, run simulations, and present methodologies, connecting GenAI tools with chip design practices.}

\subsection{Example Course 2: Build your ASIC I \& II}

\subsubsection{Build your ASIC I} The first part is a digital-design-to-silicon course experience. Students implement RTL designs, verify them with testbenches, and use the TinyTapeout~\cite{Venn24} workflows to run simulation and complete an ASIC-style flow.
The course had 12 students enrolled.

\textbf{GUIDE Units:} This course can incorporate selected GUIDE units, especially those aligned with LLM-aided design and verification. LLM-related lectures/labs can be placed after students learn basic Verilog syntax and testbench concepts.

\textbf{Semester Plan:} Table~\ref{tab:plan_build_your_asic} summarizes a typical week-by-week plan aligned with the course structure.

\textbf{Capstone Project:}
The task is \textit{Design and Implement an 8-bit Adder} from a repository or the student's own design, {using the TinyTapeout GitHub workflows}. Students  write additional test cases.
%The final submission
%includes a short report and workflow-backed artifacts: (1) explanation of the  8-bit adder design, (2) the Docs workflow PDF artifact that documents the design, (3) the Tests workflow outputs, including the waveform,
% simulation status, and screenshots, (4) the GDS workflow outputs including the GDS files and key flow results, and (5) a brief analysis of implementation statistics such as cell usage and routing. Students share Git repositories so instructors can reproduce results.
Final submission includes a design report and TinyTapeout workflow artifacts: documentation, simulation waveforms, GDS files, and implementation statistics (area/routing).

% \textbf{Assessment and deliverables:} Include weekly labs, proposal milestones, workflow-backed project progress, and project artifacts. Grading focuses on reproducible evidence produced by the tool flows, together with clear documentation of what was built and how it was validated.

\subsubsection{Build your ASIC II}
Team-based projects emphasizing C-centric \textit{high-level synthesis} (\textit{HLS}) methodologies are conducted.
Students work in teams of two and execute projects significantly more complex than the first part, demonstrating end-to-end design ability from algorithmic specification through FPGA or ASIC implementation. 
The course had 6 project teams. 
The course follows a presentation-driven milestone structure with weekly presentations. Final deliverables include: (1) weekly progress presentations, (2) final project report (5--6 pages minimum) covering design
methodology, simulation and synthesis results, FPGA/ASIC implementation, and TinyTapeout process discussion, (3) GitHub repository with C source code, generated RTL, testbenches, and implementation files, (4) YouTube
video ($\sim$10 minutes) showing the working functionality and corner cases on an FPGA board.

\subsection{Example Course 3: GUIDE4HardwareSecurity}
% This course targets students with basic digital design background and focuses on using LLMs for hardware security applications, including both attack and defense scenarios. The course spans 14 weeks and integrates foundational RTL generation concepts with specialized hardware security units. The course had 34 students enrolled, comprising undergraduate and graduate students from Electrical, Computer and Systems Engineering and Computer Science departments.
This course targets students with a basic background in digital design and focuses on using LLMs for hardware security applications, including both attack and defense scenarios. The course spans 14 weeks and integrates foundational RTL generation concepts with specialized hardware security units. Part of this course will be used at RPI for Spring 2026 by Dr. Basu for his ``Hardware Security" course.
%, where Dr. Basu will use these resources to augment his Hardware Security course. 
with \whchg{34 students enrolled}, comprising undergraduate and graduate students from Electrical, Computer and Systems Engineering and Computer Science departments. \whadd{Students work in teams (2--4 per team) for the two capstone projects, which constitute 50\% of the course grade.}

\textbf{GUIDE Units:} GUIDE4HardwareSecurity selects units from two GUIDE topics: (i) LLM-aided RTL generation (Weeks 1-4) and (ii) LLM-aided hardware security (Weeks 5-14), covering both attack and defense perspectives.

\textbf{Semester Plan:} See Table~\ref{tab:llm4hwsec_plan}. The first four weeks establish foundational skills in RTL generation and LLM-aided design workflows. Weeks 5-14 focus exclusively on hardware security applications, alternating between attack-oriented and defense-oriented units to provide comprehensive security awareness.

\textbf{Capstone:} Teams insert stealthy Trojans using LLM-based workflows (e.g., GHOST~\cite{Faruque24}) that pass regression tests but trigger under specific conditions, apply defenses (e.g., TrojanLoC~\cite{XiaoTrojanLoC25}), and evaluate attack–defense effectiveness. Deliverables are modified RTL, detection reports, LLM interaction logs, and simulation evidence. The project  has been used in class and at \textit{CSAW 2025}~\cite{csaw25}.

\subsection{Example Course 4: Hardware Design at NYU-AD)
}
This course targets students with basic digital-logic knowledge and trains them to design advanced Verilog circuits using both manual and LLM-driven approaches for efficiency comparison. 
The 14-week structure uses the first 7 weeks for core combinational/sequential design concepts and AI-vs-manual implementation practice. 
The last 7 weeks cover advanced GenAI workflows (datasets, benchmarking, RTL generation, simulation, and verification), culminating in a processor and neural-accelerator class project. 
The Spring 2026 class has 10 computer-engineering students, and enrollment is expected to grow with semiconductor initiatives in GCC/MENA region.

\textbf{GUIDE Units:} We select units from the first two topics of GUIDE, i.e., LLM-aided RTL generation, and  verification.
%, and (iii) hardware security.

\textbf{Semester Plan:} See Table~\ref{tab:hardwareDesign}. 
Weeks 1--7 cover sequential-design concepts, structural/RTL/behavioral Verilog coding, and simulation/debugging, with labs (e.g., counters, adders, traffic-light controller) implemented both manually and with LLM tools (e.g., ChatGPT, Gemini). Weeks 8--14 focus on GUIDE units for LLM-aided RTL generation (e.g., AutoChip, ROME, Veritas, VGen, VeriContaminated) and verification techniques such as testbench generation and Hybrid-NL2SVA.

\textbf{Class/Capstone Project:} 
The final project is 
\textit{LLM-Based Verilog Neural Accelerator Generation and Verification}. Students complete an end-to-end LLM-aided workflow using adders, multipliers, MAC units, and interconnects to build a systolic-array-like design. They generate MAC variants with GUIDE/commercial LLM tools, then evaluate area/performance after validation and verification using an enhanced LLM-aided testbench generator. Final submissions include a report, prompts, RTL, testbenches, and results, which can also be contributed back to GUIDE repositories.

%% file: 6-Discussion.tex
\section{Discussion and Future Directions}
\label{sec:discussion}

% GUIDE was derived from a series of tutorials offered at \textit{DATE 2024}, \textit{ETS 2024}, and \textit{ESWEEK 2024}.
% These tutorials are open-sourced on GitHub~\cite{blocklove2024llmsforeda} and are a jumping-off point for the Colabs designed for these units.

% GUIDE can support different courses by letting instructors select units and assemble them into a coherent semester offering.
% For example, \textit{LLM4ChipDesign} focuses on GenAI-assisted RTL generation and verification, including architectural analysis for LLM-aided hardware design settings and constraints.
% This course teaches students to write clear specifications, generate RTL with LLMs, and validate them with simulation-based and
% assertion-based validation. 
% \textit{Build your ASIC} centers on a design-to-silicon workflow based on TinyTapeout, emphasizing reproducible artifacts such as documentation outputs, simulation traces, GDS results, and flow statistics.
% In short, these courses reinforce different goals: \textit{LLM4ChipDesign} integrates generation with verification, while \textit{Build your ASIC} connects these to physical design via the TinyTapeout workflow.

% GUIDE also supports security-focused units. LLM-aided Hardware Trojan insertion, used in our course and in \textit{CSAW 2025}~\cite{csaw25},
% trains students to understand Trojan objectives. The project runs validation where the modified
% design passes normal regression tests; Trojan effects require targeted tests. 

GUIDE provides modular, reusable courseware for GenAI-driven digital design education, derived from tutorials~\cite{blocklove2024llmsforeda} at \textit{DATE 2024}, \textit{ETS 2024}, and \textit{ESWEEK 2024}. 
This \whchg{open-source} GitHub repository supports diverse course configurations. 
% For example, LLM4ChipDesign emphasizes GenAI-assisted RTL generation and verification with architectural analysis, while Build your ASIC focuses on design-to-silicon workflows via TinyTapeout, and LLM4HardwareSecurity integrates attack and defense perspectives.
%\textbf{Community Partnership}:
One colleague who participated in the first offering suggested introductory Colabs for standard tools (Yosys/Icarus) before tackling GenAI tasks, and using peer evaluation to improve report reproducibility. Such suggestions  strengthen the pedagogical scaffolding, by ensuring students revisit EDA concepts and enabling rigorous reporting alongside GenAI workflows.
% GUIDE is an \whchg{open-source} and collaborative repository.
% We invite researchers and educators from the EDA community to contribute new units in their areas of expertise.
We welcome contributions across all GenAI-driven design topics and encourage sharing of course modules and teaching materials.
With GUIDE becoming a community-driven resource, we aim to create a living repository that benefits educators and students worldwide. 
%We welcome contributions across GenAI-driven digital design topics and encourage sharing of course units and teaching materials deployed in classroom settings. Once GUIDE becomes a community-driven resource, it creates a living repository. % that benefits educators and students worldwide.

% GUIDE is an open and collaborative repository. We invite researchers and educators from the EDA community to contribute new units in their areas of expertise. Each contribution
% should follow the unit architecture described in Section III: slides, video, runnable Colab lab, and related papers. We welcome contributions across all GenAI-driven digital design topics and encourage sharing of course modules and teaching materials deployed in classroom settings. Once GUIDE becomes a community-driven resource, we aim to create a living repository that benefits educators and students worldwide. 
%\textbf{Expanding Coverage}: 
The repository focuses on digital design. We plan to broaden coverage to: (a) \textit{GUIDE4HLS}, studying how algorithms are translated into synthesizable C/C++ HLS code; (b) \textit{GUIDE4PhysicalDesign}, LLM-aided open-source placement and routing flows; (c) \textit{GUIDE4AnalogCircuitDesign}, LLM-aided SPICE netlist generation.
% ; and (c) \textit{GUIDE4AnalogCircuitDesign}, including LLM-aided SPICE-level netlist generation and circuit simulation.

% \textbf{Sustainability}: To keep labs runnable across semesters as LLM APIs and EDA tools evolve, we will implement lightweight continuous integration checks to detect breakage early, reduce maintenance burden, and ensure the repository remains usable as dependencies change.

\section{Acknowledgments}
The authors acknowledge the support from the Center for Secure Microelectronics Ecosystem (CSME) \#210205, NYU Center for Cybersecurity (CCS) (NYU) and CCS-NYUAD, and National Science Foundation (NSF) \#2537759, \#2347233 (NYU). Blocklove is funded in part by GAANN Fellowship. Prof. Muhammad Shafique is offering some of these modules at NYU-AD in Spring 2026 (see section IV.D).
\label{sec:acknowledgments}